\documentclass[12pt]{article}
\usepackage{graphicx}
\newcommand\beq{\begin{equation}}
\newcommand\bear{\begin{eqnarray}}
\newcommand\eeq{\end{equation}}
\newcommand\eear{\end{eqnarray}}
\begin{document}
\baselineskip=24pt

\begin{center}
{\Large \bf  First examples of stable transition metal complexes of an 
all-metal antiaromatic molecule (Al$_4$Li$_4$)} 
\end{center}

\vspace*{0.5cm}

\centerline{\bf Ayan Datta and Swapan K Pati$^*$}

\vspace*{0.1cm}

\begin{center}
Theoretical Sciences Unit and Chemistry and Physics of Materials Unit, \\
Jawaharlal Nehru Center for Advanced Scientific Research  \\
Jakkur Campus, Bangalore 560 064, India. E-mail: pati@jncasr.ac.in
\end{center}

\begin{center}
{\bf Abstract}
\end{center}

\vspace*{0.2cm}

We propose new methodologies for stabilizing all-metal antiaromatic clusters 
like: Al$_{4}$Li$_{4}$. We demonstrate that these all-metal species can be stabilized 
by complexation with 3d-transition metals very similar to its organic counterpart, 
C$_4$H$_4$. Complexation to transition metal ions reduce the frontier orbital 
energies and introduces aromaticity. We consider a series of 
such complexes [$\eta$$^4$(Al$_4$Li$_4$)-Fe(CO)$_3$, $\eta$$^2$$\sigma$$^2$(Al$_4$Li$_4$)-Ni 
and (Al$_{4}$Li$_{4}$)$_{2}$Ni] and make a comparison between the all-metal species and the 
organometallic compounds to prove conclusively our theory. Fragmentation energy 
analysis as well as NICS support similar mechanism of complexation induced 
stability in these all-metal molecules.  

\newpage
\clearpage

The concept of aromaticity and antiaromaticity is of fundamental importance in 
chemistry. From the simple H${\ddot{u}}$ckel theory to the more refined 
concepts like 
ring-currents and critical point charge-densities, the field has evolved over a 
period of half-a-century\cite{aroma}. The idea has been extended from initially 
a small class of organic $\pi$-conjugated systems to now many inorganic molecules 
and molecular clusters\cite{inorg}, with rapid experimental verifications
through actual synthesis and characterizations\cite{inorg1}.

The recent report of the first all-metal antiaromatic complex, Al$_4$Li$_4$ 
shows the generalizations and usefulness 
of this idea\cite{antiaroma}. Also, in the last few years there have been 
reports of aromaticity in all-metal clusters\cite{antiaroma1}. However, 
unlike their organic counterpart, C$_4$H$_4$, 
where the energy separation between the $\sigma$ and $\pi$ orbitals 
is substantial, these all-metal molecules have closely placed orbitals and 
thus have poor $\sigma$-$\pi$ separation. Mostly, the aromatic characteristic 
is associated with delocalized electrons ($\pi$ electrons). However 
recently it has been reported that these clusters are more $\sigma$-aromatic 
than $\pi$-antiaromatic\cite{schleyer}. As a result, there has been a confusion 
whether to call these complexes aromatic or antiaromatic. The controversy can 
only be settled after the successful synthesis followed by unambiguous 
crystal structure determinations. For a stable molecular crystal, 
measurement of bond lengths alternations as well as the charge 
densities (ring critical points) are well established parameters for 
characterizing aromaticity/antiaromaticity\cite{crystal}. 

These clusters however have till now been synthesized only in 
the gas-phase by laser 
vaporization technique and is thus insufficient in providing 
structural details. 
Recently, we have shown that these materials, Al$_4$M$_4$ (M=Li, Na and K) 
are also very good candidates for higher-order nonlinear optical (NLO) 
applications due to charge transfer from the highly electropositive ion (Li) 
to the Al$_4$-rings\cite{ayan}. Unfortunately, the stability is one issue 
which hinders any applications let alone firm establishment of the basic 
understanding.

The synthesis of antiaromatic molecules is difficult because of their
instabilities. Cyclobutadiene, (C$_4$H$_4$), a 4$\pi$ electron system 
remained non-isolated for a longtime before Longuet-Higgins and 
Orgel proposed in a landmark paper, the concept of stabilization 
through complexation with a transition metal to form an organometallic 
compound\cite{higgins}. The compound was 
synthesized soon after\cite{crie}. In the following, we justify this 
simplistic model for the small Al$_4$-clusters and propose a few very stable complexes 
of these all-metal species. Parallelly, we also compare and contrast the
energetics with their organic analogues (C$_4$H$_4$ complexes).

We have performed a closed shell calculation for 
singlet and an open shell calculation for the triplet state at the 6-311G(d,p)
basis set level. Electron correlation has been included according to the DFT method
using Becke's three parameter hybrid formalism and the Lee-Yang-Parr functionals
(B3LYP) available in the Gaussian electronic structure set of 
codes\cite{becke}. The geometries obtained from the B3LYP method have been shown 
to be in very good agreement with the measured photoelectron spectra in such 
small clusters\cite{supp}. 

In Fig. 1, the minimum energy ground state structures are shown for C$_4$H$_4$ 
and Al$_4$Li$_4$. 
Simple H${\ddot{u}}$ckel $\pi$-electron theory predicts a triplet square 
geometry for C$_4$H$_4$ with equal C-C bond lengths\cite{salem}. However
inclusion of interaction with the underlying $\sigma$ backbone
stabilizes the C$_4$H$_4$ molecule in a singlet state with rectangular
geometry. This is a good example of Jahn-Teller distortion or Pierls 
instability in low-dimensional system which allows stabilizations 
through bond length alternation. In fact, in this picture, the square geometry 
actually corresponds to a transition state between two degenerate
rectangular ground state structures.
In Table 1, the total energies, bond length alternation 
($\Delta r$, defined as the average difference between the 
bond lengths of two consecutive bonds in the 4-membered ring) 
for both the states are tabulated. 
The rectangular C$_4$H$_4$ 
[A(i)] is more stable than the square geometry [A(ii)] by 6.2 
Kcal/mol. Thus, a triplet square geometry is expected to be the 
transition state for processes such as ring whizzing, where one 
rectangular form is converted into the other (an in-plane rotation 
of 90$^0$), in harmony with time-resolved transition state 
studies for the tub-inversion in 1,3,5,7-cyclooctatetraene\cite{zewail}.

Since in Al$_4$Li$_4$ the $\sigma$-$\pi$ separation is poor, 
the H${\ddot{u}}$ckel $\pi$-electron picture is completely invalid. In fact,
the $\pi$ electrons in this case interacts more strongly with the
$\sigma$ backbone and we expect a distorted structure as the 
ground state. The ground 
state structure for the singlet state is found to be a rectangular 
Al$_4$-geometry with surrounding Li atoms, forming a C$_{2h}$ symmetry group 
[B(i)]. The same structure has been found in previous calculations 
as well\cite{schleyer}. Another low-energy structure
for Al$_4$Li$_4$ found by optimizing the geometry for the singlet square
geometry is a diamond shaped structure. It has a D$_{2h}$ symmetry [B(ii)]
and is 15 Kcal/mol higher in energy than the stable C$_{2h}$ geometry [B(i)].
Thus, the rectangular Al$_4$-ring in [B(i)] corresponds to the ground-state
geometry for Al$_4$Li$_4$. The structural distortion leading to a magnetic triplet
state with D$_{4h}$ symmetry [B(iii)] is found to lie 55 Kcal/mole above the
ground state singlet [B(i)]. There also exist a low-energy structure with
the same geometry as the ground state (C$_{2h}$ symmetry) with purely 
spin excitation at an energy only 5 Kcal/mole above the ground state. This
triplet geometry [B(iv)] for Al$_4$Li$_4$ does not have a counterpart in 
C$_4$H$_4$, clearly explaining the low $\sigma$-$\pi$ gap and the existance 
of continuum like metallic states in Al$_4$Li$_4$.  

Existence of a very stable rectangular ground state structure together with a 
square geometry as the transition state for the Al$_4$-ring similar to 
those for C$_4$H$_4$ suggests that these inorganic clusters are 
anti-aromatic in nature. This is expected since, the highly 
electropositive Li atoms donate electrons to the Al-atoms, thereby 
creating a species of the type Al$_4$$^{4-}$, isoelectronic with 
C$_4$H$_4$. Thus we can safely consider Al$_4$Li$_4$ as a 4$\pi$ electron 
system with the $\pi$-HOMO (highest occupied molecular orbital) being a non-bonding 
molecular orbital as like in C$_4$H$_4$. In the context of C$_4$H$_4$, 
Longuet-Higgins 
suggested that such a system can be stabilized if the non-bonding electrons 
form bonding molecular orbitals with suitable low energy d-orbitals of a 
transition metal. For this to happen however, the energies of the d-orbitals
should lie close to the low-energy levels of the molecule alone. In the
following, we propose a few stable complexes of Al$_4$Li$_4$ and compare their
formation energies in comparison with their organic analogues.

\section{{\bf Fe(CO)$_3$ complex}:} A molecular 
complex, $\eta$$^4$(C$_4$H$_4$)-Fe(CO)$_3$, has been recognized 
through the formation of such bonding molecular orbitals and this complex
has been reported to be quite stable\cite{pettit1}. In fact, oxidation of this complex 
releases the C$_4$H$_4$ ligand which is a stable source for the highly reactive 
cyclobutadiene in organic synthesis\cite{pettit2}. For the Al$_4$Li$_4$, 
we perform ground 
state energy analysis on the similar system, $\eta$$^4$(Al$_4$Li$_4$)-
Fe(CO)$_3$, using the same level of theory mentioned above. Both 
$\eta$$^4$(Al$_4$Li$_4$)-Fe(CO)$_3$ and its organic analogue have substantial 
stability (see Fig. 2 for structures). Al$_4$Li$_4$ indeed forms a 
stable $\eta$$^4$ complex with Fe(CO)$_3$ [2(ii)]. The stability of the 
complexes are investigated using the following fragmentation scheme:

\begin{center}
$\eta$$^4$(C$_4$H$_4$)-Fe(CO)$_3$ = C$_4$H$_4$ + Fe(CO)$_3$
\end{center}
\begin{center}
$\eta$$^4$(Al$_4$Li$_4$)-Fe(CO)$_3$ = Al$_4$Li$_4$ + Fe(CO)$_3$
\end{center}

The binding energy for $\eta$$^4$(Al$_4$Li$_4$)-Fe(CO)$_3$ 
is found to be 106.04 Kcal/mol while that for $\eta$$^4$(C$_4$H$_4$)-
Fe(CO)$_3$ it is 78.44 Kcal/mol. The comparable binding energies for the 
two compounds suggest that Al$_4$Li$_4$ is very well stabilized in the complex, 
in-fact even more stabilized than C$_4$H$_4$.

The HOMO of the molecule (a non-bonding MO for 
C$_4$H$_4$), interacts with the low-energy d-orbital of the ligand to form 
a bonding combination in the complex. Thus, C$_4$H$_4$ that initially 
possessed 4$\pi$ electrons now 
has two more electrons, forming a species of the type 
C$_4$H$_4$$^{2-}$, an aromatic molecule. Similarly, for Al$_4$Li$_4$, the
complexation converts it into Al$_4$Li$_4$$^{2-}$, a well-established aromatic 
complex\cite{antiaroma1}. The HOMO energies for C$_4$H$_4$ in the free and in the 
coordinated form (derived by performing a single-point energy calculation 
on the C$_4$H$_4$ fragment in the optimized complex) are $-0.198$ au 
and $-0.157$ au respectively, 
while the same for the complex, $\eta$$^4$(C$_4$H$_4$)-Fe(CO)$_3$, is
$-0.250$ au. Similarly, for Al$_4$Li$_4$, the free and the coordinated 
form have HOMO energies at $-0.128$ and $-0.104$ respectively and the complex 
$\eta$$^4$(Al$_4$Li$_4$)-Fe(CO)$_3$ has the same at
$-0.168$ au. The stabilization of the frontier orbitals in the metal complex 
in both the systems has its manifestation at 
the formation of the stable structure. The similarity in the difference between
the HOMO energies of free and coordinated structures and with that
of the corresponding complex for each molecule suggest that a similar 
mechanism is operative in lowering the energies of the frontier orbitals
in stabilizing the complexes.

The above explanation is based on a simplistic picture of the interaction. To verify 
that indeed such a scheme is valid for a molecule with poor $\sigma$-$\pi$ 
separation like Al$_4$Li$_4$, we compare the $\Delta r$ for both C$_4$H$_4$ 
and Al$_4$Li$_4$ in the free geometry and when they are complexed with the 
transition metal. For C$_4$H$_4$, the $\Delta r$ is 0.24 $\AA$ in the 
free state. In the complex, $\eta$$^4$(C$_4$H$_4$)-Fe(CO)$_3$, the $\Delta r$ 
for the C$_4$H$_4$ ring is only 0.005 $\AA$. Thus, C$_4$H$_4$ when complexed is a 
square rather than a rectangle and as expected from the $\pi$-only interaction, it 
behaves as aromatic C$_4$H$_4$$^{2-}$. For Al$_4$Li$_4$, $\Delta r$=0.13 $\AA$ 
in the free state while in the complex it is only 0.03 $\AA$. This clearly supports 
that the Al$_4$Li$_4$ has been converted into Al$_4$Li$_4$$^{2-}$, accounting for 
its substantial stability due to aromaticity. The complexation 
induced metalloaromaticity in Al$_4$Li$_4$ is schematically 
shown in Fig. 3. While a square (triplet) Al$_4$Li$_4$ is much higher in 
energy than the rectangular Al$_4$Li$_4$, this square form is stabilized on 
complexation to a transition metal. Same is the case for C$_4$H$_4$, where the 
square form becomes stabilized upon complexation. This is similar to the origin
of aromaticity in benzene, where the $\pi$-delocalized D$_{6h}$ structure 
corresponds to a 
energy minima between two bond-altered Kekule forms with D$_{3h}$ symmetry.

An even more clear picture is derived by performing a calculation for the 
nucleus-independent chemical shift (NICS)\cite{scheleyer1} at the GIAO-B3LYP
/6-311+G(d,p) level. We calculate the NICS at the center of the Al$_4$ ring 
before and after complexation with the Fe(CO)$_3$. For comparison, the 
same values are also calculated for C$_4$H$_4$.
In C$_4$H$_4$, NICS values 
before (C$_4$H$_4$) and after complexation (C$_4$H$_4$$^{2-}$) are 23.55 ppm 
and -15.37 ppm 
respectively. The change in sign clearly shows the transition from 
antiaromatic to aromatic nature upon complexation. For Al$_4$Li$_4$, the NICS
values change from -11.01 ppm 
in the free state to -25.44 ppm on complexation in $\eta$$^4$(Al$_4$Li$_4$)-
Fe(CO)$_3$. The initial negative magnitude for NICS in free Al$_4$Li$_4$ 
supports the claim by Schleyer et al that Al$_4$Li$_4$ has higher 
$\sigma$-aromaticity than $\pi$-antiaromaticity\cite{schleyer}. But, the 
increase in NICS value with same negative sign suggests increased aromaticity 
in these clusters upon complexation, which is expected from the $\pi$ only picture 
of the conversion of Al$_4$Li$_4$ to Al$_4$Li$_4$$^{2-}$. Thus, complexation 
with Fe(CO)$_3$ induces metalloaromaticity in Al$_4$Li$_4$ and thereby 
stabilizes the complex, $\eta$$^4$(Al$_4$Li$_4$)-Fe(CO)$_3$. 

\section{{\bf bis nickel(II) chloride complex}:} Next we consider another very well known 
example of a stable C$_4$H$_4$ 
complex, bis(cyclobutadiene nickel(II) chloride). The tetramethyl derivative 
for the complex crystallizes in a P21/c point group and has a good resolution
(R=7.0$\%$), CCD reference code, NCBNIB \cite{dunitz}. We have obtained the 
structure from the database and the methyl groups was substituted by H for 
easy comparison with the Al$_4$Li$_4$ derivative, bis(Al$_4$Li$_4$ nickel
(II) chloride). Both the structures, bis(cyclobutadiene nickel(II) chloride) 
and bis(Al$_4$Li$_4$ nickel(II) chloride) were optimized at the same level of 
theory as mentioned above. The structure for bis(cyclobutadiene nickel(II) chloride) 
remains similar to that found from the crystal structure. Fig. 4 shows 
the structures for the two complexes. For the organometallic 
complex 4(i), the bond length alternation in the C$_4$H$_4$ ring is only 
0.05 $\AA$. Therefore this bridged chlorine system also shows strong
mixing of the d-orbitals from Ni and the non-bonding electrons of 
C$_4$H$_4$. 

For the all-metal complex 4(ii) however, the $\eta$$^4$(Al$_4$Li$_4$)-Ni 
binding mode is converted into $\eta$$^2$$\sigma$$^2$(Al$_4$Li$_4$)-Ni
upon optimization. In Fig.5, we show this change in the binding mode. Two of 
the initial $\pi$-bonds between Al and Ni (Al-Ni distance of the order 
2.42 $\AA$) are now converted into strong $\sigma$ bonds (Al-Ni 
distance of the order 2.25 $\AA$) in the optimized geometry. Such a change in 
bonding pattern from $\pi$ character to $\sigma$ character is quite well known in 
organometallic chemistry\cite{ometal}. It is interesting to note that a similar 
phenomenon occurs in the all metal complexes as well.

A fragmentation analysis on these two molecules is given below.

\begin{center}
bis(cyclobutadiene nickel(II) chloride) = 2 C$_4$H$_4$ + bis(nickel(II) chloride)
\end{center}
\begin{center}
bis(Al$_4$Li$_4$ nickel(II) chloride) = 2 Al$_4$Li$_4$ + bis(nickel(II) chloride)
\end{center}

The stabilization energies are 68.39 Kcal/mol for 4(i) and 286.77 Kcal/mol for 4(ii). 
Such high stability in the Al$_4$Li$_4$ complex is due to the formation of 
two strong Al-Ni $\sigma$ bonds as discussed above.

\section{{\bf Metal sandwich complex}:} Another well known methodology in 
stabilizing a molecule is to form 
a sandwich type of geometry where two molecular species can share interaction 
with a transition metal: cyclopentadiene is stabilized in 
such a geometry resulting in the Ferrocene structure\cite{wilkin}. For 
C$_{4}$H$_{4}$, a simple effective 
electron number (EAN) counting shows that the metal in 
between the two ligands should have 10 valence electrons in stabilizing
a sandwich of the type:(C$_{4}$H$_{4}$)$_{2}$M. The simplest metal
with 10 electrons in the valence shell is Nickel(0). Elements in the same group 
like Pd or Pt have a strong spin-orbit coupling and prefer square-planar geometry 
(16 electron geometry). Thus a coordination number of 8 as required 
in a sandwich complex is not possible with Pd or Pt. After performing the geometry
optimization at the same level of theory discussed above, we find that the 
structure for (C$_{4}$H$_{4}$)$_{2}$Ni is indeed a sandwich geometry with the two 
C$_{4}$H$_{4}$ rings above and below the Ni atom (see Fig. 6 (i)). In this
complex, the Ni atom sits symmetrically inside the cavity of the two C$_{4}$H$_{4}$ 
rings with a distance of 1.99$\AA$ from each C$_4$H$_4$ ring. The two C$_{4}$H$_{4}$ 
are staggered to each other.  

Similarly, we have been able to stabilize the Al$_{4}$Li$_{4}$ cluster 
by introducing 
it in a sandwich of the type: (Al$_{4}$Li$_{4}$)$_{2}$Ni. 
The geometry is shown in Fig. 6 (ii) (the optimization is performed at the
B3LYP/LANL2MB level 
followed by energy calculation at B3LYP/6-311G(d,p) level). The central Ni atom 
sits unsymmetrically in the cavity of the two Al$_{4}$Li$_{4}$ rings. A very 
recent theoretical study on its aromatic analogue, Al$_{4}$$^{2-}$, support our 
calculations\cite{mercero}. Interestingly, the Al atoms in the 
rings bend towards the Ni atom and the planarity of the Al$_{4}$ ring is thereby 
lost. This is understood from the fact that when the 4$\pi$ electrons 
of each of the two Al$_{4}$Li$_{4}$ rings interact with the central Ni atom,
the requirement of the Al atoms to be in plane with the Li atom is no
longer important. Instead, the sandwich like structure with 18
electrons gives an extra stabilization keeping the whole system
electrically neutral. The stability of these complexes are investigated 
using the following fragmentation scheme:

\begin{center}
(C$_{4}$H$_{4}$)$_{2}$Ni = 2 C$_{4}$H$_{4}$ + Ni(0)
\end{center}
\begin{center}
(Al$_{4}$Li$_{4}$)$_{2}$Ni = 2 Al$_{4}$Li$_{4}$ + Ni(0)
\end{center} 
where Ni(0) is in a $^3$F state. Al$_{4}$Li$_{4}$ binds strongly to the Ni(0) and 
has a binding energy of 134.285 Kcal/mol. For C$_{4}$H$_{4}$ this binding energy is 
150.819 Kcal/mol.

In conclusion, we have demonstrated for the first time that the all-metal species like 
Al$_{4}$Li$_{4}$ can be stabilized by complexation with 3d-transition metals, very
similar to its organic counterpart, C$_4$H$_4$. Although such a complexation 
induced metalloaromaticity 
is a well established concept in the realm of organometallic chemistry, we have 
demonstrated that it is a very general and elegant concept which can be used for 
all metallic molecules with orbitals 
that are close in energy with the d-orbitals of the transition metal. 
We have also shown that these all-metal complexes have similar binding energies 
and properties like their organometallic counterparts and thus should be considered 
as very good candidates for experimental synthesis. Also, 
such a stabilization will provide a very precise answer to the question of 
aromaticity/antiaromaticity and will lead to novel applications of these clusters.
We believe that our work will motivate 
synthesis of these molecules as were the case in the previous generation for 
the organometallic complexes.

Acknowledgements: 
We would like to dedicate this communication to the memory of Prof. H. C. Longuet-Higgins. 
SKP acknowledges CSIR and DST, Govt. of India, 
for the research grants.

\pagebreak
\clearpage

\newpage

\begin{table}
\caption{The total energies (in au) and the bond-length alternation, $\Delta r$ 
(in $\AA$) 
for C$_{4}$H$_{4}$ and Al$_{4}$Li$_{4}$ in different spin states corresponding to
different lowest energy structures.}
\end{table}

\begin{center}
\begin{tabular}{|l|l|l|l|l|}
\hline
Molecule  & Symmetry & Spin-state & Energy & $\Delta r$    \\ \hline
C$_{4}$H$_{4}$(A(i))   & D$_{2h}$ &  Singlet   & -154.718  & 0.240        \\ \hline
C$_{4}$H$_{4}$(A(ii))   & D$_{4h}$ &  Triplet   & -154.708  & 0.000       \\ \hline
Al$_{4}$Li$_{4}$(B(i)) & C$_{2h}$ &  Singlet   & -999.932  & 0.130        \\ \hline
Al$_{4}$Li$_{4}$(B(ii)) & D$_{2h}$ &  Singlet   & -999.908  & 0.120       \\ \hline
Al$_{4}$Li$_{4}$(B(iii)) & D$_{4h}$ &  Triplet   & -999.844  & 0.000       \\ \hline
Al$_{4}$Li$_{4}$(B(iv)) & C$_{2h}$ &  Triplet   & -999.926  & 0.200       \\ \hline
\end{tabular}
\end{center}

\begin{figure}
\includegraphics[scale=0.65] {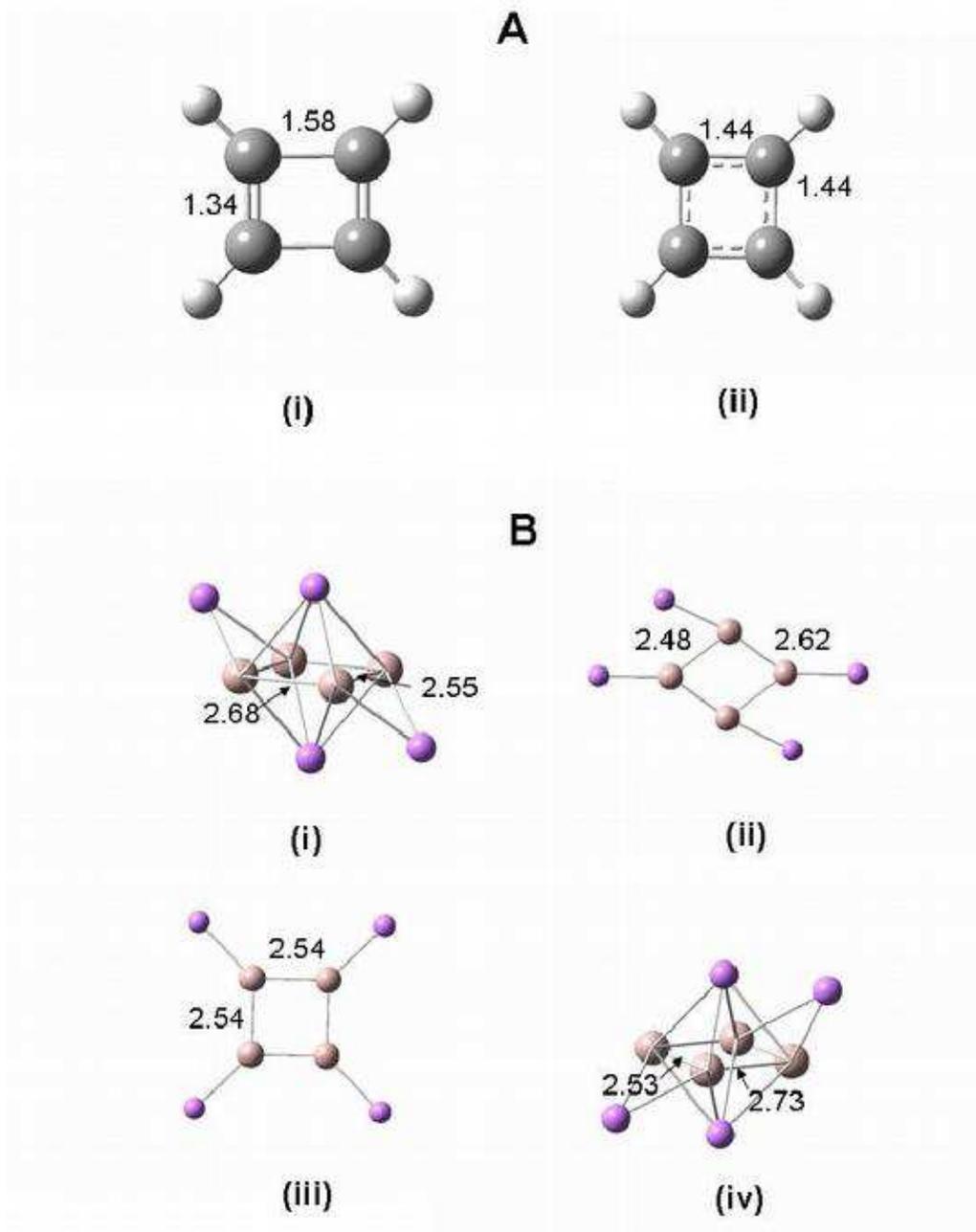}
\caption{Equilibrium minimum energy geometries for C$_{4}$H$_{4}$ and
Al$_{4}$Li$_{4}$ in the singlet and triplet states (See Table. 1). 
Bond lengths are given in $\AA$.
Ball color: Black=C, White=H, Pink=Li, Light orange=Al.}
\end{figure}

\begin{figure}
\includegraphics[scale=0.65] {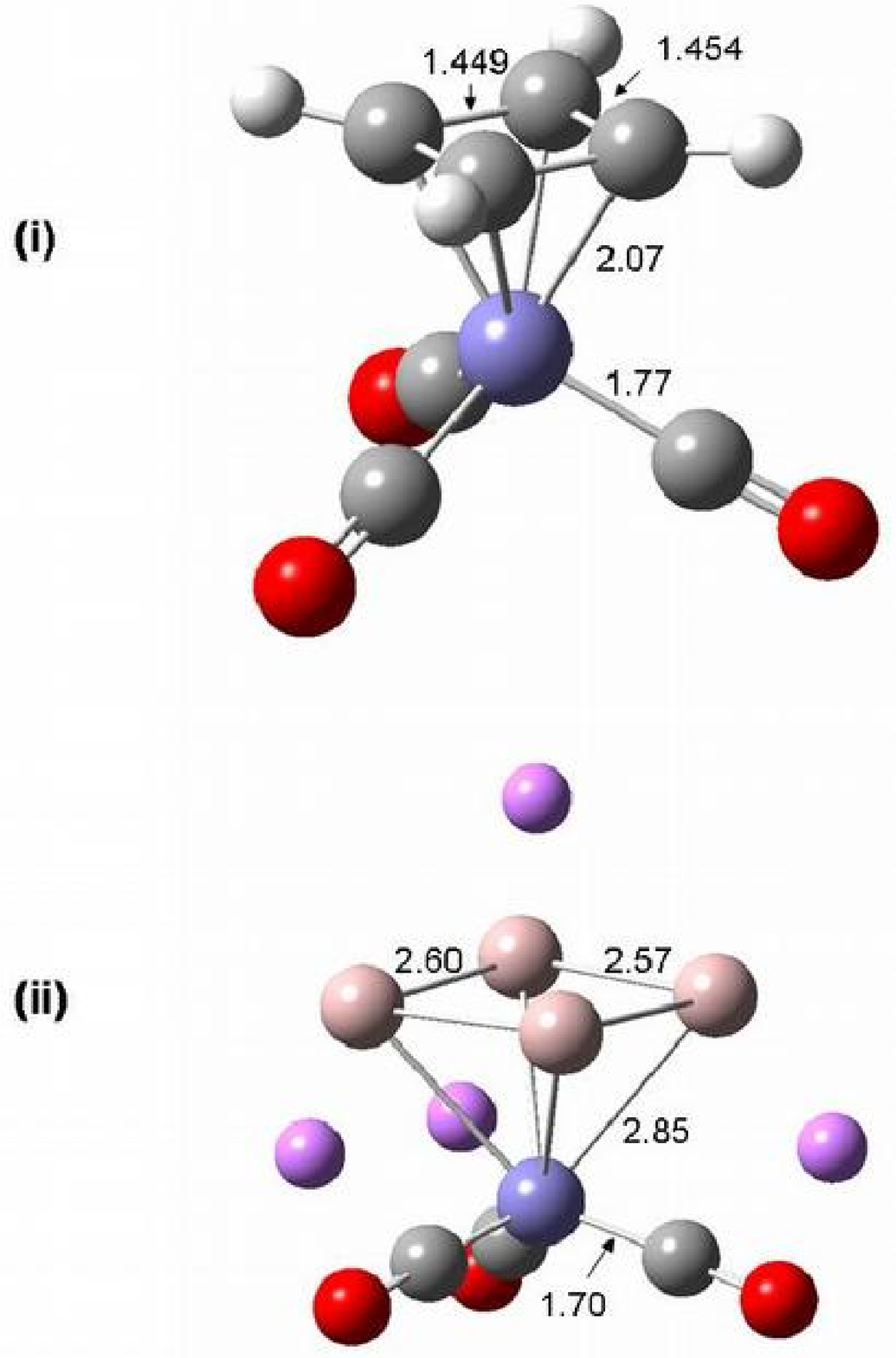}
\caption{Equilibrium minimum energy geometries for (i) $\eta$$^4$(C$_4$H$_4$)
-Fe(CO)$_3$ and (ii) $\eta$$^4$(Al$_4$Li$_4$)-Fe(CO)$_3$. 
Bond lengths are in $\AA$.
Ball color: Red=O, Violet=Fe.}
\end{figure}

\begin{figure}
\includegraphics[scale=0.6] {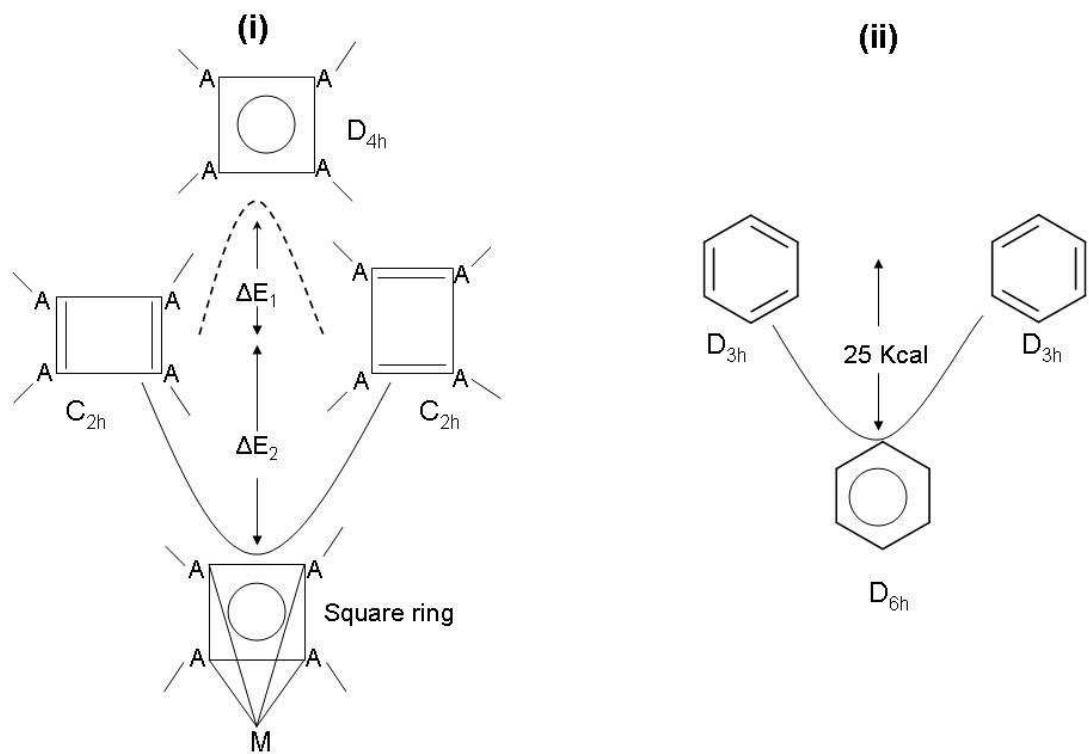}
\caption{Schematic representation of (i) change in geometry for ring 
whizzing and complexation to transition metal center for A=Al in Al$_4$Li$_4$ 
($\Delta E$$_1$=55 Kcal/mol, $\Delta E$$_2$=100 Kcal/mol); A=C in C$_4$H$_4$ 
($\Delta E$$_1$=6.2 Kcal/mol, $\Delta E$$_2$=78.4 Kcal/mol) 
(ii) Ring whizzing in benzene.}
\end{figure}

\begin{figure}
\includegraphics[scale=0.7] {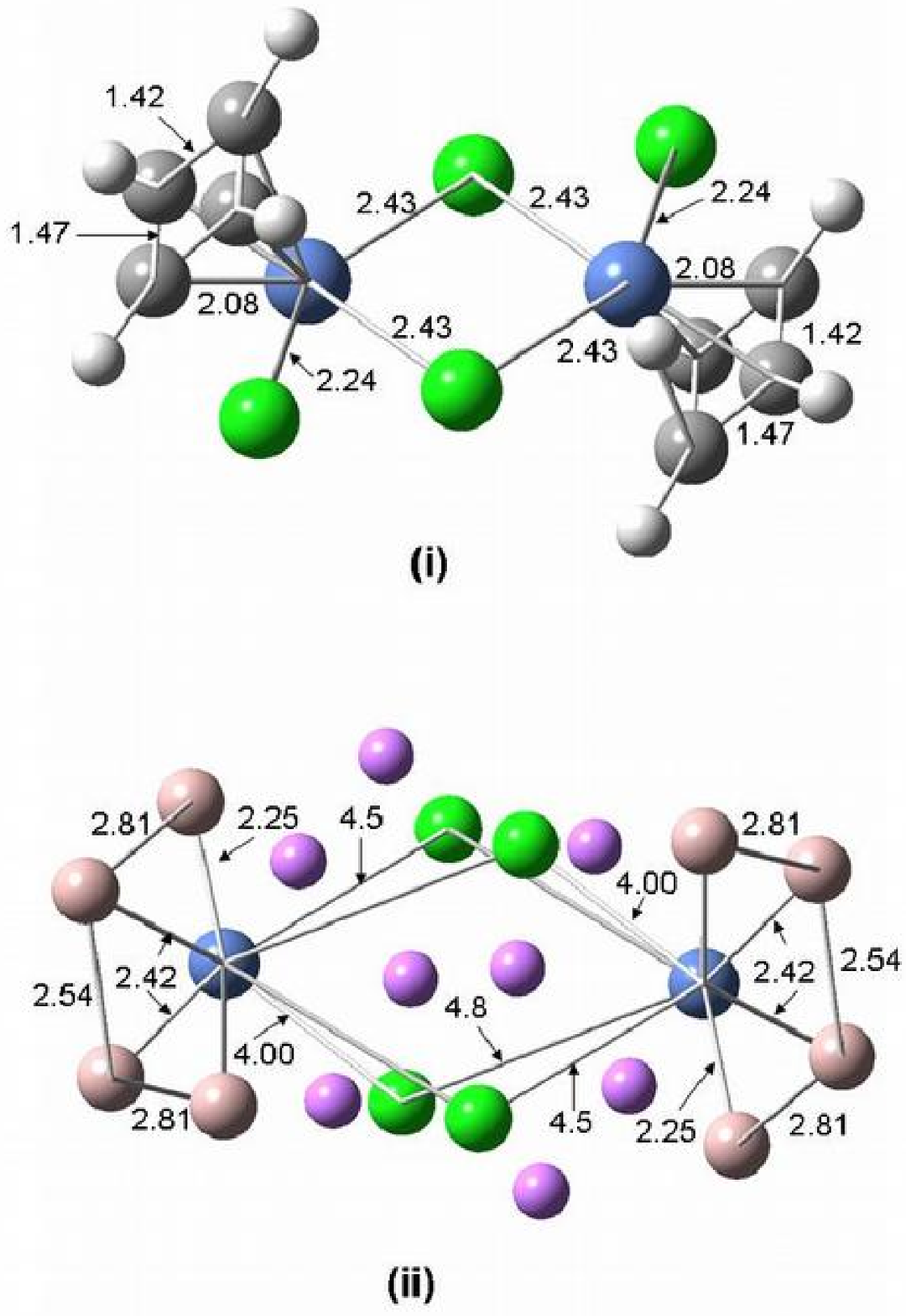}
\caption{Equilibrium minimum energy geometries for (i) bis(cyclobutadiene
nickel(II) chloride)and (ii) bis(Al$_4$Li$_4$ nickel(II) chloride). Distances 
are in $\AA$. Ball color: Green=Cl, Blue=Ni.}
\end{figure}

\begin{figure}
\includegraphics[scale=0.7] {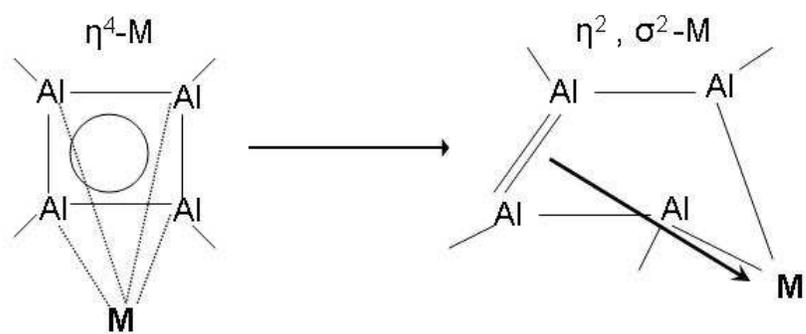}
\caption{Conversion from $\eta$$^4$ binding mode for Al$_4$Li$_4$ to $\eta$$^2$-
$\sigma$$^2$ in the complex in Fig. 4(ii).}
\end{figure}

\begin{figure}
\includegraphics[scale=0.65] {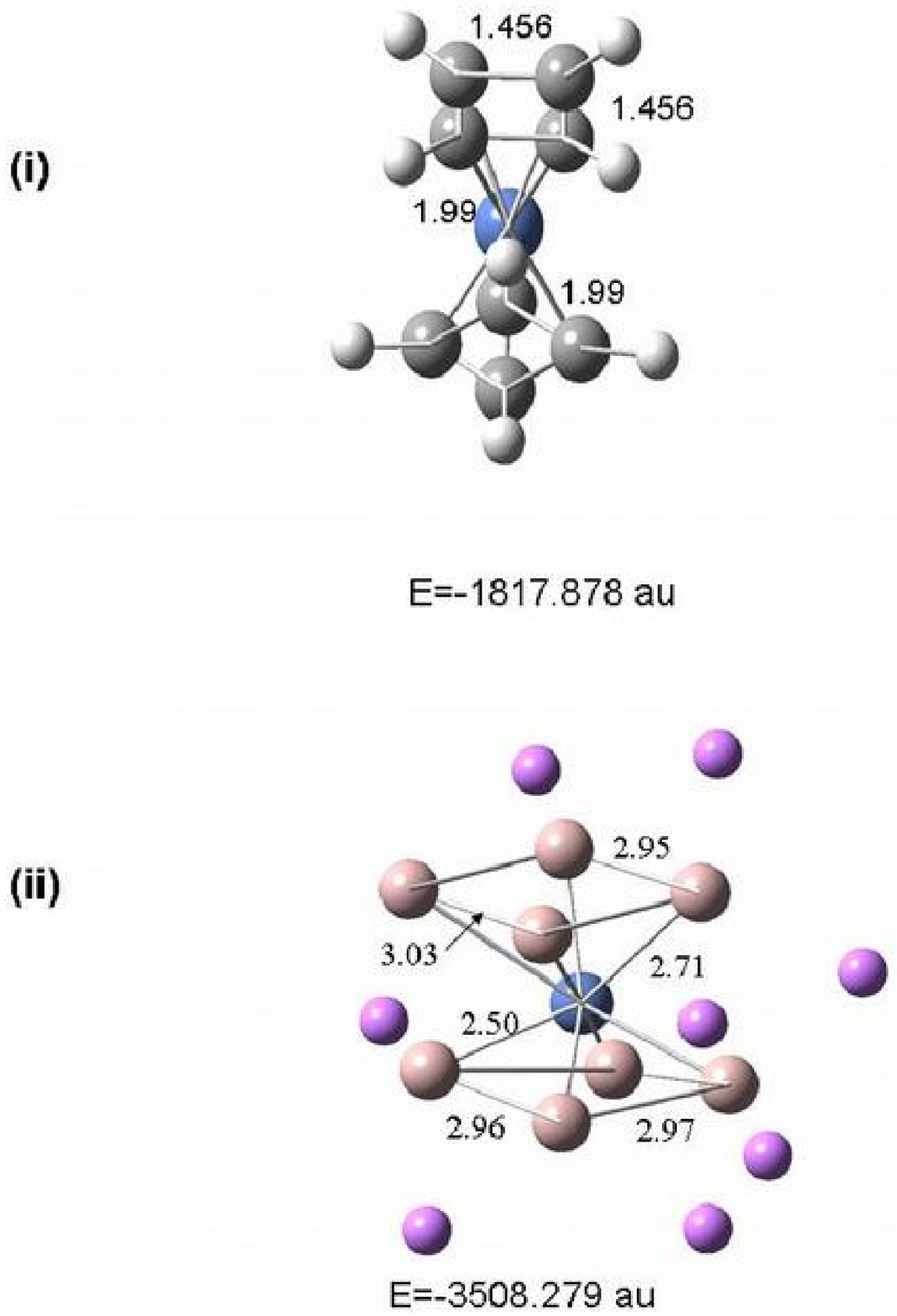}
\caption{Equilibrium minimum energy geometries for (i) (C$_{4}$H$_{4}$)$_{2}$Ni and 
(ii) (Al$_{4}$Li$_{4}$)$_{2}$Ni. Distances are in $\AA$.}
\end{figure}

\end{document}